\documentclass[12pt]{article}
\usepackage[T2A]{fontenc}
\usepackage[english]{babel}
\usepackage[english]{babel}
\usepackage[dvips]{graphicx}
\usepackage{mathtext}

\begin{document}

\title{Space number density of bright quasars \\ in the halo model of galaxy formation}
\author{Kulinich~Yu., Novosyadlyj~B.}
\maketitle

\medskip

\centerline{\it Astronomical Observatory of Ivan Franko National University of Lviv}
\medskip

\begin{abstract}
We analyse the redshift dependence of space number density of quasars assuming that they are the short-lived active stages of the massive galaxies and arise immediately after the collapse of homogeneous central part of protogalaxy clouds. Obtained dependence fits the observational data ChaMP+CDF+ROSAT (Silverman et al. 2005) very well for protogalaxy clouds of mass $M\approx 8\cdot 10^{11}$~$h^{-1}M_{\odot}$ and ellipticity $e<0.4$. The lifetime of bright X-ray AGNs or QSOs with $L_X>10^{44.5}$~erg$\cdot s^{-1}$ in the range of energies $0.3-8$~keV is  $\tau_{QSO}\sim 6\cdot10^6$ years when the mass of supermassive black hole is $M_{SMBH}\sim 10^{9}$~$M_{\odot}$ and the values of other quasar parameters
are reasonable. The analysis and all calculations were carried out in the framework of $\Lambda$CDM-model with parameters determined from 5-years WMAP, SNIa and large scale structure data (Komatsu et al. 2009). It is concluded, that the halo model of galaxy formation in the $\Lambda$CDM cosmological model matches well observational data on AGNs and QSOs number density coming from current optical and X-ray surveys. 


PACS number(s): 98.54.Aj, 98.54.Kt, 98.54.-h, 98.65.Fz, 98.35.Jk

\end{abstract} 

\section*{Introduction}
The radio, optical and X-ray observational data indicate that co-moving space number density of bright AGNs and QSOs is unmonotonous function of redshift: it increases  up to $z\sim 2.5$, declines and goes to zero after that \cite{Boyle88,Fan01,Hawkins96,Shaver99,Schmidt95,Silverman05}. Small angular diameters and huge luminosities of QSOs signify that their central power engines are supermassive black holes (SMBH) which efficiently transform the mass of accreting matter into radiation. About ten percents of mass can be converted into the high energy radiation by this mechanism instead of decimal parts of percent for the nucleosynthesis reactions providing the luminosity of stars. Observations using Hubble space telescope and the largest ground-based ones have revealed the host galaxies of QSOs in which they occupy central parts. High emissivity of quasars, their popularity in comparison with galaxies, genetic closeness to galaxy active nuclei support the idea that QSOs are short-lived active phases of galaxies with SMBH  at their nuclei. So, the space number density of quasars can be described in the scenario of galaxy formation taking into account the peculiarities which make them QSOs. Really, in such approach the redshift distribution of QSOs space number density becomes qualitatively understood: the growth of QSOs number density in the redshift range $z=(\infty,\,\,\sim 2.5]$ is caused by the rate of galactical nuclei formation, the later decay in the redshift range $z=[\sim 2.5,\,0]$  is due to the short-lived action of quasar engine (finite fuel resource in the vicinity of galaxy centrum and the large rate of its burning) as well as to the decreasing of galaxy birth-rate. In this paper we develop the semi-analytical approach proposed in the previous papers \cite{Cen92,Nuser93,Haehnelt93,Novosyadlyj97,Chornij04} in order to adjust quantitatively the prediction of concordance cosmological $\Lambda$CDM-model \cite{Komatsu09} to observational data on bright QSOs number density given by \cite{Silverman05}.

\section{Formation of bright QSO{\small s} in the halo model}
Dependence of the QSOs number density on redshift is determined by several factors: birth-rate of galaxies of relevant mass, probabilities of SMBH formation in their centra and  the tempo of accretion of matter to them. The birth-rate of galaxies of given mass is determined completely by cosmological model and initial power spectrum of density perturbations. For the estimation of other factors the additional assumptions are needed. The high popularity of quasars suggests that the formation of SMBH is quite frequent phenomenon. The physical conditions in the centrum of forming galaxy are favourable for formation of the central SMBHs through the collapse of dark matter and hydrogen-helium gas \cite{Haehnelt93,Umemura93,Loeb94,Eisenstein95,Bromm03,Koushiappas04,Begelman06} as well as through the merging of remnants of supermassive stars \cite{Abel00,Kauffmann00,Bromm02,Volonteri03}. In both cases the time-lag between halo virialization and the birth of the bright quasar is expected to be short comparing to the cosmological time-scale, even at high redshifts. For appearance of the luminous quasar the one solar mass of gas per year must infall into SMBH. A few physical mechanisms of gas supplying to quasar engine are analysed in literature. For example, \cite{Menci03} suppose that accreation of gas into SMBH is caused by the tide collisions of close galaxies in the process of their merging. The semi-analytical models of the growth of SMBHs in the central parts of their host galaxies, developed by \cite{Malbon07,Marulli08,Somerville08,Bonoli09}, fit well the observational data on quasar luminosity function and two-point space correlation one 
for selected redshifts.

In this paper we assume that bright QSOs appear in the process of galaxies formation and their luminous efficiency is provided by the collapsing gas from nearest vicinity of SMBH. The durations of both processes are essentially lower than cosmological time. So, quasars are supposed to be the short-lived active phases of the galaxies formation, which is well described by the halo model. The central density peak of protogalaxy cloud corresponding to the bottom of gravitational potential well is practically homogeneous and collapses first of all with formation of SMBH. The outer shells will collapse later and the relation between their co-moving radius $R$ and the collapse time $t_{col}$ is given by simple formula following from the model of spherical collapse \cite{Eke96,Kulinich08}: 
$$\frac{M(R)-\overline{M}_R}{\overline{M}_R}D(1) = \delta_c(t_{col}),$$
where $M(R)=4\pi\int\limits_0^R\rho(r)r^2dr$ is the mass of the matter in the sphere with co-moving radius R, $\rho(r)=\overline{\rho}(1+\delta(r))$, $\overline{\rho}$ is the mean matter density at moment $t_{col}$, $\overline{M}_R \equiv \frac43\pi R^3\overline{\rho}$, $\delta$ is the initial amplitude of density perturbation, $\delta_c(t_{col})$ is the critical overdensity as function of collapse time, $D(a)$ is growth factor to the present from high redshift relative to the growth factor from the same initial amplitude and initial redshift in the Einstein-de Sitter model, $a$ is the scale factor, which equals 1 at current epoch. This formula is correct if mass of matter inside each shell is fixed. For close to  centrum shells this condition is satisfied at the initial stage of SMBH formation because particles belonging to them have small target parameters. Outer shells with lower averaged density contrast will collapse later when hydrodynamical and dark matter halo virialization processes take place. Some part of matter will fall into a trap of SMBH providing the quasar luminous efficiency. The quasar will be active some time after formation of galaxy and virialization of dark matter halo till the matter which serve the fuel for quasar engine will be depleted from the range of unstable orbits. The tide collision of merging galaxies can renew 
the quasar activity and this can be repeated many times during the galaxy life. We suppose that most luminous quasars are short-lived active stages of early evolution of galaxies. 

For calculation of bright QSOs number density at different redshifts we have used the analytical approach proposed by \cite{Press74} and improved by \cite{Lacey93}.
Let us suppose that masses of protogalactic clouds (protohalos) which will host brightest QSOs are in the range $M,\,M+\Delta M$ and masses of homogeneous central parts of perturbations (top-hat)
in which the SMBH will be formed are in the range $M_{th},\, M_{th}+\Delta M_{th}$. The mass of SMBH can be lower than $M_{th}$. 

The halo number density with mass in the range $M_{th},\, M_{th}+\Delta M_{th}$ at some time $t$ is determined by the mass function \cite{Press74}
\begin{eqnarray*}
 n(M_{th},t) = \frac{\overline{\rho}}{M_{th}}\frac{\delta_c}{(2\pi)^{1/2}\sigma^3_{th}}
\exp\left[-\frac{\delta_c^2}{2\sigma^2_{th}}\right]\frac{d\sigma^2_{th}}{dM_{th}}\Delta M_{th},
\end{eqnarray*}
where $\sigma^2_{th}\equiv \sigma^2(M_{th})$ is the r.m.s. of density perturbations in the top-hat sphere with radius $R_{th}=(3M_{th}/4\pi\overline{\rho})^{1/3}$ and $\delta_c(t)$ is the critical magnitude of the linear density perturbation which collapses at $t$. 
We will take into account only that haloes with mass $M_{th}$ which are central peaks of protogalaxy clouds with mass $M$. They can be extracted from the total number density by estimation of the probability that its mass after collapse and virialization of protogalaxy cloud will be in the range $M,\,M+\Delta M$ during short time $\Delta t$ \cite{Lacey93}
\begin{eqnarray*}
 \frac{d^2p}{dMdt}(M_{th}\to M|t)\Delta M\Delta t = \frac{1}{(2\pi)^{1/2}}\left[\frac{\sigma^2_{th}}{\sigma^2(\sigma^2_{th}-\sigma^2)}\right]^{3/2}\times
\exp\left[-\frac{\delta_c^2(\sigma^2_{th}-\sigma^2)}{2\sigma^2_{th}\sigma^2}\right]\frac{d\sigma^2}{dM}\frac{d\delta_c}{dt}\Delta M\Delta t,
\end{eqnarray*}
where $\sigma^2\equiv \sigma^2(M)$ is the r.m.s. of density perturbations in the sphere which contains the mass $M$ of the whole halo.

The important condition for SMBH formation is spherical symmetry and top-hat density profile of central region of protogalaxy cloud in which the appearance of bright quasar is expected. All shells in such region collapse practically simultaneously. This should result in formation of SMBH. As the main fraction of mass consists of
the collisionless cold dark matter particles, the pressure of hot baryonic gas can not prevent the collapse. So, from estimated number density of haloes with given mass we must extract the essentially non-spherical ones. The distribution of the ellipticity $e$ and prolateness $p$ of ellipsoidal clouds with given initial amplitude of density perturbation $\delta$ has been obtained by \cite{Bond96,Sheth01}: 
\begin{equation}
 g(e,p|\delta) = \frac{1125}{\sqrt{10\pi}}e(e^2-p^2)\left(\frac{\delta}{\sigma}\right)^5e^{-2.5(\delta/\sigma)^2(3e^2+p^2)},\label{gep}
\end{equation}
where $0\le e < \infty$ and $-e\le p \le e$. We suppose that ellipticity of homogeneous regions of protogalactic clouds, the collapse of which leads to SMBH formation, is constrained from above: $e\le e'_m$. Their fraction is as follows
 \begin{equation}
 p(e<e'_m) = \int\limits_0^{e'_m}de\int\limits_{-e}^edp g(e,p|\delta).
\end{equation}
This integral has functional presentation in the form $p(e<e'_m) \equiv f(x)$, where $x = e'_m\delta/\sigma(M_{th})$. It means that constraining of ellipticity of central homogeneous regions $e\le e'_m$ constrains the ellipticity of the whole protogalactic cloud $e\le e_m$, where $e'_m/\sigma(M_{th})=e_m/\sigma(M)$.

Finally, we present the QSOs number density in the co-moving space as product of the number density of haloes with mass in the range $M_{th},\,M_{th}+\Delta M_{th}$, the fraction of those of them which belong to larger haloes with mass in the range $M,\,M+\Delta M$ and the fraction of such haloes with ellipticity lower than $e_m$:
\begin{eqnarray}
 n_{QSO}(t) \simeq \frac{\overline{\rho}}{M_{th}}\frac{f\left(e_m\delta_c/\sigma\right)}{4\pi{\left[\sigma^2(\sigma^2_{th}-\sigma^2)\right]}^{3/2}}\frac{d\delta^2_c}{dt}
\times\exp\left[-\frac{\delta_c^2}{2\sigma^2}\right]\frac{d\sigma^2}{dM}\frac{d\sigma^2_{th}}{dM_{th}}\Delta M_{th}\Delta M\Delta t.
\label{nqsot}
\end{eqnarray}
We present the dependence of QSOs number density on redshift in the form:
\begin{eqnarray}
 n_{QSO}(z) = A \cdot f\left(\frac{e_m\delta_c(z)}{\sigma(M)}\right)
\frac{d\delta^2_c(z)}{dz}\left(\frac{1}{H_0}\left|\frac{dz}{dt}\right|\right)
\times\exp\left[-\frac12\left(\frac{\delta_c(z)}{\sigma(M)}\right)^2\right],\label{nqsoz}
\end{eqnarray}
where 
\begin{equation}
 \frac{1}{H_0}\frac{dz}{dt}=-(1+z)^{2.5}\sqrt{\Omega_m + \frac{\Omega_K}{1+z} + \frac{\Omega_{\Lambda}}{(1+z)^3}}.
\end{equation}
All redshift-independent multipliers in (\ref{nqsot}) are collected by $A$ ($[A]=$Mpc$^{-3}$). It is the product of relation $\frac{\overline{\rho}}{M_{th}}$, some constants, spectral dependent values ($\sigma,\;\;\sigma_{th}$), their derivatives with respect to mass as well as of the unknown values such as the QSO lifetime ($\tau_{QSO}=\Delta t$) and mass range of haloes ($\Delta M$) which maintain the bright QSOs or X-ray AGNs. Therefore, we assume it is a free parameter of model. The critical amplitude of density perturbations $\delta_c(t_{col})$ depends on matter density and value of cosmological constant \cite{Kulinich03,Kulinich08}. For $\Lambda$CDM-model with parameters used here it equals 1.673 at $z=0$. For other redshifts we have approximated the numerical results for $\delta_c(z)$ from \cite{Kulinich03} by the simple analytic formula 
$$\delta_c(z)=\left(\frac{0.564}{0.39+(1+z)^{2.86}}+1.267\right)\left(1+z\right),$$ 
which works well in the range of redshifts $0\le z\le 5$. So, the dependence of QSOs number density (\ref{nqsoz}) on redshift has 3 free parameters: the normalization constant $A$, mass $M$ and maximal ellipticity of the protogalactic clouds $e_m$. They are used for the fitting of $n_{QSO}(z)$ to the observational data on QSOs number density.

\section{Results and discussions}
\begin{figure}
\centerline{\includegraphics[width=8.0cm]{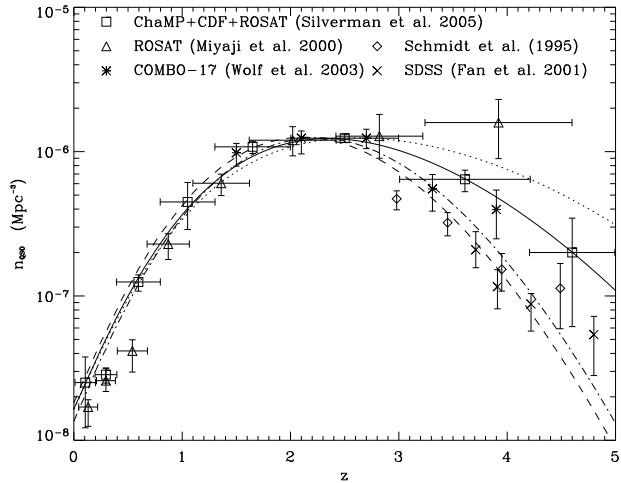}}
\caption{The dependence of QSOs co-moving number density on redshift: theoretical curves versus observational data. Solid line shows the best fitting model with $M=8\cdot 10^{11}$~$h^{-1}M_{\odot}$ and $e_m=0.38$. Dashed and dash-dotted lines present the models with the same protogalactic clouds masses $M=4\cdot 10^{12}$~$h^{-1}M_{\odot}$, but different ellipticities, $e_m=0.2$ and $e_m=0.17$ respectively. Dotted line shows the co-moving space number density of QSOs for the model with $M=3\cdot 10^{11}$~$h^{-1}M_{\odot}$ and $e_m=0.5$.}
\label{grape}
\end{figure}

We have used the observational data on bright QSOs number densities at different redshifts presented by \cite{Silverman05}, which are the compilation of optical (SDSS, COMBO-17) and X-ray (Chandra, ROSAT) QSOs and AGNs redshift surveys (see Fig.\ref{grape}). The ChaMP+CDF+ROSAT data, which include the X-ray AGN with $L_X>10^{44.5}$~erg$\cdot s^{-1}$ in the energy range $0.3-8$~keV, are shown in the figure by squares.
The results of calculations of $n_{QSO}(z)$ for $\Lambda$CDM-model with parameters  $\Omega_b=0.044$, $\Omega_m = 0.258$, $\Omega_{\Lambda}=0.742$, $h=0.719$, $\sigma_8 = 0.796$ and $n_s=0.963$ \cite{Komatsu09} are shown there by lines. We calculated the r.m.s. of density perturbation as follows: 
$\sigma(M)=\mathcal{A}(2\pi)^{-3/2}\int_0^\infty k^{(2+n_s)}T^2(k)W^2(Rk)dk$,
where $R$ is the radius of the top-hat sphere containing the mass $M$, $W(x)=3(\sin{x}-x\cos{x})/x^3$ is the window function and $T(k)$ is the transfer function. The latter
has been calculated using the publicly available code CAMB \cite{Lewis00}.  The normalization constant of power spectrum $\mathcal{A}$ we determined from the equality $\sigma_8^2=\mathcal{A}(2\pi)^{-3/2}\int_0^\infty k^{(2+n_s)}T^2(k)W^2(8k)dk$.

The solid line in Fig.\ref{grape} shows $n_{QSO}(z)$ calculated for $M=8\cdot 10^{11}$~$h^{-1}M_{\odot}$ and $e_m=0.38$. The parameter $A$ has been used for fitting of $n_{QSO}(z)$ to observational data at the maximum ($z\simeq 2.5$) and for this line it equals $9.75\cdot 10^{-7}$ Mpc$^{-3}$.  It is the product of lifetime of QSOs, spectral dependent values $\sigma(M),\;\;\sigma(M_{th}$) and mass range of haloes $\Delta M$ which maintain bright QSOs. So, in the $\Lambda$CDM-model the halo of mass $M=8\cdot 10^{11}$~$h^{-1}M_{\odot}$  with top-hat central part 
$M_{th}=10^{9}$~$h^{-1}M_{\odot}$ may have SMBH with $M_{SMBH}\sim 0.1-0.001 M_{th}$ which should shine like bright QSO during $\tau_{QSO}\sim 5.6\cdot 10^6$ years 
($\Delta\log{M}=\Delta\log{M_{th}}=1$). As we can see the solid line agrees very well with the observational data ChaMP+CDF+ROSAT \cite{Silverman05} covering the wide range of redshifts from $z\simeq0$ to $z\simeq5$.

Without constraint of ellipticity of  protogalaxy clouds hosting QSOs the solid line in Fig. \ref{grape} should be essentially higher than corresponding observational data at $z<1.2$. As it follows from (\ref{gep}) the mean value of ellipticity of cloud which collapse at given $z$  is  $e=\sigma(M)/(2\delta_c(z))$. For high $z$ the ellipticity is small as $\delta_c(z)$ is large. The clouds which collapse at lower $z$ are more elliptical and more rarely produce the QSOs.

The mass $M$ of protogalaxy clouds, which can host bright QSOs, depends on $\sigma_8$ and $\delta_c(z)$ and should be different in the cosmological models with different sets of parameters.  For higher  $\sigma_8$ the mass of protogalaxy will be higher.  Increasing of the mass leads to the steeper curve of QSOs number density at  $z>2.5$ and to the displacement of its maximum to lower   $z$. On the other hand, the decreasing of the upper limit of  ellipticity of protogalaxy clouds leads to displacement of  that maximum to higher  $z$. The dashed and dash-dotted lines in Fig.\ref{grape} show the QSOs number density for ellipticities $e\le e_m=0.2$ ($A=7\cdot 10^{-7}$  Mpc$^{-3}$) and $e\le e_m=0.17$ ($A=1.1\cdot 10^{-6}$  Mpc$^{-3}$) correspondingly. The mass of  protogalaxy cloud is  $M=4\cdot 10^{12}$~$h^{-1}M_{\odot}$ for both cases.

The minimal mass of halo in which the bright QSOs may arise can be estimated by comparison of the model curve for quasar number density with corresponding observational data.
Indeed, the number density of haloes with lower mass is higher and they collapse earlier. So, for  large $z$ the predicted number density curve will be less steep and its maximum will be shifted to $z>2.5$. The dotted line shows the redshift dependence of number density of bright quasars which arise in the halo with mass $M=3\cdot10^{11}$~$h^{-1}M_{\odot}$ and initial ellipticity lower than $e_m=0.5$ ($A=5.35\cdot 10^{-8}$  Mpc$^{-3}$). It lays above all observational data (excluding one ROSAT point \cite{Miyaji00}) and that mass can serve as rough estimation of lower limit for mass of haloes which produce the bright QSOs and/or X-ray AGN.

The dependence of bright QSOs number density on redshift obtained here matches the observational data essentially better than ones deduced in the frame of semi-analytical merging models of SMBHs formation (see for comparison Fig.~9 from \cite{Kauffmann00}, Fig.~3 from \cite{Malbon07}, Fig.~7 from \cite{Marulli08}, Fig.~3 from \cite{Bonoli09}), which underestimate the number density of luminous AGNs/QSOs at high redshifts. 

We must note that number density of quasars with lower luminosity,  $L_{2-8 keV}<10^{44}$~erg$\cdot s^{-1}$, has maximum at $z\sim 1$ \cite{Barger03,Cowie03,Fiore03,Steffen03, Ueda03}. This can not be explained by simple model for bright quasars used here. Some reasons should be taken into account for this case. 1) The assumption that lifetime of bright QSOs is small in comparison with cosmological time can be incorrect for fainter quasars. It can depend on density profile of protogalaxy cloud, angular momentum etc. 2) SMBHs can be formed in the protogalaxies of smaller mass essentially later then collapse has occured. The time-lag can depend on physical parameters of protogalaxy cloud too. 3) Fainter quasars can be the result of reactivation of SMBH in the centra of some galaxies by tidal collision of satellite galaxies or their merging. Some galactic nuclei can experience such reactivations several times during the galaxy life. 

\section{Conclusions} 
We have assumed that the bright QSOs are short-lived active phases of early evolution of massive galaxies. Their large X-ray and optical luminosities are provided by accretion of baryonic matter to SMBHs which have been formed  as result of collapse of homogeneous central parts of protogalaxy clouds (haloes). Using the expression for co-moving number density of haloes of mass $M_{th}$ present at time $t$ \cite{Lacey93} and constraining the ellipticity ($e\le e_m$) of those of them which belong to larger haloes of mass $M$ we have obtained the formula for dependence of QSOs number density on redshift. It has 3 free parameters ($M$, $e_m$ and normalization constant $A$) which allow us to fit the model dependence to observational data. As background cosmological model we have used the $\Lambda$CDM-model with best fitting parameters determined by \cite{Komatsu09}: $\Omega_m = 0.258$, $\Omega_{\Lambda}=0.742$, $h=0.719$, $\sigma_8 = 0.796$ and $n_s=0.963$. Obtained dependence for $n_{QSO}(z)$ (\ref{nqsoz}) fits in the best way the observational data on bright QSOs and X-ray AGN number densities ChaMP+CDF+ROSAT \cite{Silverman05} for $M=8\cdot 10^{11}$~$h^{-1}M_{\odot}$ and $e_m=0.38$. The best fitting value of normalization constant is $A=9.75\cdot 10^{-7}$  Mpc$^{-3}$. The lifetime of such QSOs $\tau_{QSO}\approx 6\cdot 10^6$ years. The lower limit of halo mass estimated in such approach is $M=3\cdot 10^{11}$~$h^{-1}M_{\odot}$.

Therefore, the halo model of galaxy formation together with the assumption that bright QSOs are short-lived active phases of early evolution of massive galaxies quantitatively agrees well with corresponding observational data. 
 
\section*{Acknowledgments}
This work was supported by the projects of Ministry of Education and Science of Ukraine ``Investigation of variable stars, supernova remnants and stellar clusters using observational data obtained by groundbased and space telescopes'' and ``Formation of large-scale structure in the Universe with dark energy''. The authors appreciate a partial support from the  National Academy of Sciences of Ukraine under the research program ``Cosmomicrophysics''.

\label{lastpage}
\end{document}